\documentclass[conference]{IEEEtran}
\IEEEoverridecommandlockouts
\usepackage{cite}
\usepackage{amsmath,amssymb,amsfonts}
\usepackage{algorithmic}
\usepackage{graphicx}
\usepackage{textcomp}
\usepackage{xcolor}
\def\BibTeX{{\rm B\kern-.05em{\sc i\kern-.025em b}\kern-.08em
    T\kern-.1667em\lower.7ex\hbox{E}\kern-.125emX}}

\usepackage{graphicx}
\usepackage{latexsym}
\usepackage{amsmath}
\newtheorem{definition}{Definition}
\usepackage{booktabs} 
\usepackage{multirow}
\usepackage{booktabs}
\usepackage{graphicx}  
\usepackage{diagbox}
\usepackage[braket]{qcircuit} 
\usepackage{threeparttable} 
\usepackage{stfloats} 
\usepackage{enumitem} 
\usepackage{fancyheadings} 

\usepackage{makecell}
\usepackage{times}
\usepackage{graphicx}
\usepackage{epsf}
\usepackage{verbatim}
\usepackage{url}
\usepackage{color}
\usepackage{alltt}

\newcommand{\Comment}[1]{}

\usepackage{braket} 
\usepackage{listings}
\usepackage{color}
\usepackage{graphicx} 
\usepackage{float} 
\usepackage{subfigure}
\usepackage[]{hyperref}

\lstdefinelanguage{QSharp}
{morekeywords={operation, Int, if, body, Unit, let, for, in, Controlled, adjoint, Adj, Ctl, is, Qubit},
sensitive=true,
morecomment=[l]{//},
}
\begin{document}

\title{ScaffML: A Quantum Behavioral Interface Specification Language for Scaffold}

\author{\IEEEauthorblockN{Tiancheng Jin}
\IEEEauthorblockA{\textit{Kyushu University} \\
Fukuoka, Japan \\
jintc1@f.ait.kyushu-u.ac.jp
}
\and
\IEEEauthorblockN{Jianjun Zhao}
\IEEEauthorblockA{\textit{Kyushu University} \\
Fukuoka, Japan \\
zhao@ait.kyushu-u.ac.jp
}
}

\maketitle

\begin{abstract}
Ensuring the correctness of quantum programs is crucial for quantum software quality assurance. Although various effective verification methods exist for classical programs, they cannot be applied to quantum programs due to the fundamental differences in their execution logic, such as quantum superposition and entanglement. This calls for new methods to verify the correctness of quantum programs. In this paper, we present a behavioral interface specification language (BISL) called ScaffML for the quantum programming language Scaffold. ScaffML allows the specification of pre- and post-conditions for Scaffold modules and enables the mixing of assertions with Scaffold code, thereby facilitating debugging and verification of quantum programs. This paper discusses the goals and overall approach of ScaffML and describes the basic features of the language through examples. ScaffML provides an easy-to-use specification language for quantum programmers, supporting static analysis, run-time checking, and formal verification of Scaffold programs. Finally, we present several instances to illustrate the workflow and functionalities of ScaffML.
\end{abstract}


\section{Introduction}\label{sec:intro}

Model-oriented specification languages draw inspiration from two influential papers by Hoare \cite{hoare1969axiomatic,hoare1978proof}. The first paper \cite{hoare1969axiomatic} introduced the concept of pre- and postconditions, which are predicates over program states that specify the desired behavior of a computation. The precondition describes the state requirements before the computation, while the postcondition describes the expected final state. The second paper \cite{hoare1978proof} introduced the notion of abstraction functions, which map the implementation data structures to a mathematical value space. This allows the specification of abstract data types (ADTs), enabling reasoning about the ADT operations without needing to consider the implementation details. Model-oriented specification languages leverage these ideas to specify software modules such as procedures, functions, and methods using pre- and postconditions. These specifications utilize a vocabulary defined in an abstract model, which mathematically defines the abstract values and their properties.

The two widely used model-oriented specification languages are VDM~\cite{jones1990systematic} and Z~\cite{spivey1992z}. These languages offer a mathematical toolkit that allows users to construct abstract models for specifying procedures. The VDM toolkit resembles that of a (functional) programming language, providing fundamental types such as integers, booleans, and characters, as well as structured types like records, Cartesian products, disjoint unions, and sets. On the other hand, the Z toolkit is based on set theory and offers a rich notation for various set constructions, along with powerful techniques for combining specifications (schema calculus).

Generic model-oriented specification languages like Z and VDM have the capability to specify programs across different programming languages. However, they may not fully capture the precise interfaces of modules specific to a particular programming language. This limitation arises due to the variations in interface details among different programming languages. To address this, behavioral interface specification languages (BISLs) have been developed, specifically tailored to individual programming languages~\cite{hatcliff2012behavioral}. Examples of such BISLs include ACSL for C~\cite{baudin2008acsl} and JML~\cite{leavens2006preliminary} for Java. The advantage of using a BISL designed for a specific programming language is the ability to specify both the behavior and the precise interface of the program. Recent research has demonstrated that employing BISLs for specifying programs in a specific programming language yields significant practical benefits in terms of static compile-time checking and run-time checking~\cite{hatcliff2012behavioral}.

Research on formal specification languages needs to adapt to the emergence of new language paradigms to specify programs written in these new languages effectively. This requires developing new specification approaches specifically tailored to these languages, including quantum programming languages.

Quantum programming has gained significant attention in recent years as it involves designing and implementing executable quantum computer programs to achieve specific computational tasks. Various quantum programming languages, such as Scaffold~\cite{abhari2012scaffold}, Qiskit~\cite{aleksandrowicz2019qiskit}, Q\#~\cite{svore2018q}, and Quipper~\cite{green2013quipper}, have been developed to facilitate the writing of quantum programs. With the growing maturity of quantum programming research and the availability of active research tools, it is crucial for researchers in the field of formal specification languages and BISLs, to address the unique challenges posed by this new paradigm.

The field of quantum programming has predominantly focused on problem analysis, language design, and implementation, leaving the specification and validation of quantum programs with limited attention. In order to formally verify quantum programs, it is crucial to have a means of formally specifying their properties. However, despite the existence of numerous generic formal specification languages and BISLs for classical procedural and object-oriented programming languages, no BISL has been developed, to the best of our knowledge, specifically for specifying programs written in quantum programming languages like Scaffold. Furthermore, the unique features of quantum programming languages, such as quantum superposition, entanglement, and no-cloning, pose challenges in directly applying existing formal specification languages and BISLs designed for classical programming languages. Therefore, we are motivated to design a formal specification language tailored for expressing properties of programs written in quantum programming languages.

Rather than aiming for a generic specification language, we have chosen to develop a BISL specifically for Scaffold, a quantum programming language~\cite{abhari2012scaffold}. A BISL allows us to describe the interface details and behavior of modules from the client's perspective, providing a formal specification of both the behavior and the exact interface of Scaffold programs' modules. This step is essential for formally verifying these modules.

Our BISL for Scaffold is named ScaffML (Scaffold Modeling Language), which follows similar approaches as classical BISLs like ACSL~\cite{baudin2008acsl} for C and JML for Java~\cite{baudin2008acsl,leavens2006preliminary}. ScaffML introduces annotations for specifying Scaffold modules and interfaces, including pre- and postconditions and assertions. These annotations facilitate dynamic analysis for debugging and testing, as well as static analysis for formal verification of the properties of Scaffold programs. Static analysis verifies that a Scaffold module's code correctly implements its specification. To enable the verification of quantum programs, we are developing a language conversion tool that automatically transforms Scaffold programs with ScaffML specifications into an intermediate language suitable for theorem provers such as Coq~\cite{Formalproof} or CVC3~\cite{barrett2007cvc3}. This conversion process allows for the formal checking and verification of Scaffold programs.

In this paper, we discuss the goals of ScaffML and its overall specification approach, providing examples of how to use ScaffML to specify Scaffold modules and programs.

The rest of the paper is organized as follows. Section~\ref{sec:rationale} presents the design rationale for ScaffML, highlighting the motivation behind its development. In Section~\ref{sec:background}, we briefly introduce The ANSI/ISO C Specification Language (ACSL), which serves as the foundation for ScaffML. Section~\ref{sec:ScaffML} demonstrates the practical application of ScaffML through examples, showcasing how Scaffold modules and programs can be effectively specified using the language. Section~\ref{sec:tool-support} discusses the crucial aspect of tool support for ScaffML. 
Section~\ref{sec:related-work} discusses related work. Finally, Section~\ref{sec:conclusion} concludes the paper and outlines future directions for ScaffML.


\section{Design Rationale}\label{sec:rationale}

Our primary objective in developing ScaffML is to investigate the formal specification and verification of quantum programs. The design of ScaffML represents just one aspect of our approach. In addition to creating ScaffML, we are committed to developing techniques and tools that facilitate the formal specification and verification of Scaffold programs integrated with ScaffML specifications. To achieve this, we have intentionally designed ScaffML as a compatible extension of Scaffold. This design choice serves two primary purposes: (1) to simplify the adoption of ScaffML by existing Scaffold users and (2) to enable the utilization of the well-established ACSL-based verification toolchain for verifying Scaffold programs.

ACSL serves as a particularly suitable foundation for the design of ScaffML due to two key reasons. Firstly, Scaffold, being an extension of C for quantum programming, aligns well with ACSL, which is a dedicated BISL crafted specifically for C. By structuring ScaffML as an extension of ACSL, we can direct our focus towards the unique challenges arising from the utilization of quantum modules. Secondly, ACSL offers a robust toolchain that efficiently supports both static and dynamic analysis of C programs. Thus, if we envision the automatic transformation of Scaffold programs with ScaffML specifications into corresponding verification conditions (VC) using VC generators, these conditions can subsequently be fed to interactive provers or automatic provers such as Coq and CVC3. Importantly, due to the similarity in approaches between ScaffML and ACSL, we can leverage the ACSL toolchain for the verification of Scaffold programs.

Similar to ACSL, ScaffML specifications are expressed as comments within Scaffold interface definitions, denoted by two forward slashes (//). This design allows ScaffML specifications to be seamlessly integrated into regular Scaffold files.

\section{ANSI/ISO C Specification Language} \label{sec:background}
ACSL (ANSI/ISO C Specification Language)\cite{baudin2008acsl} is a BISL specifically designed for the C programming language\cite{kernighan1988c}. It enables the specification of preconditions, postconditions, and assertions for C functions. Predicates in ACSL are expressed using regular C expressions augmented with logical operators and universal and existential quantifiers. ACSL specifications are written as special comments within C function definitions, enclosed between \verb+/*@+ and \verb+*/+.

ACSL supports the specification of C functions at both the function and statement levels. These two levels of specification form the complete behavioral specifications for the function. At the function level, the specification consists of requirements over the function's arguments and properties that must hold upon completion. The requirements are expressed as preconditions, while the properties ensured at the end of the function are called postconditions. This specification forms a contract between the function and its callers: callers must ensure that the precondition holds before invoking the function, and in return, the function guarantees the postcondition upon its return.

ACSL also supports three types of statement-level specifications: \textit{assertions}, \textit{loop invariants}, and \textit{statement contracts}. Assert statements can be placed before any C statement or at the end of blocks and represent properties that must hold at specific program points. Loop invariants can be specified before loop statements such as \verb+while+, \verb+for+, and \verb+do while+, providing hints to static analyzers for accurate loop analysis. Statement contracts are allowed before any C statement, including blocks, and describe the behavior of the statement, akin to function specifications.

This paper adopts ACSL as the foundation to develop ScaffML for specifying Scaffold programs. To emphasize the core concepts of ScaffML, we do not address the specification of Scaffold classical modules in this paper, as they can be specified using ACSL in a similar manner~\cite{baudin2008acsl} for C.

\section{Specifying Quantum Gates in ScaffML}\label{sec:Gates}
The execution of a quantum program involves the application of various quantum gates. Ensuring the correctness of these operations is crucial for the functionality of the quantum program. In this section, we present examples of how to specify certain gates in ScaffML, leveraging the standard quantum gate libraries provided in Scaffold.

\subsection{Quantum Bits and Quantum Gates}

Quantum bits, commonly referred to as \textit{qubits}, are fundamental elements in quantum programs. While classical bits have defined values of 0 or 1, qubits can exist in a linear superposition of these two basis states. The basis states $\ket{0}$ and $\ket{1}$ can be represented as $\ket{0}$=[1, 0]$^\top$ and $\ket{1}$=[0, 1]$^\top$, respectively. A qubit's quantum state can be expressed as $\ket{q}$ = $\alpha\ket{0}$ + $\beta\ket{1}$, where $\alpha$ and $\beta$ are complex numbers satisfying $\left| \alpha \right|^2+\left| \beta \right|^2=1$. These complex numbers, known as probability amplitudes, determine the behavior of qubits.

In quantum computing, quantum gates play a pivotal role in manipulating the state of qubits within a quantum circuit. Applying a quantum gate to a qubit can alter its state. It is important to note that quantum gates are reversible, meaning the number of inputs and outputs must always be the same. Quantum gates serve as the fundamental building blocks of quantum circuits and are indispensable for performing quantum computations.

\begin{figure}[h]
\centerline{\includegraphics[width=0.95\linewidth]{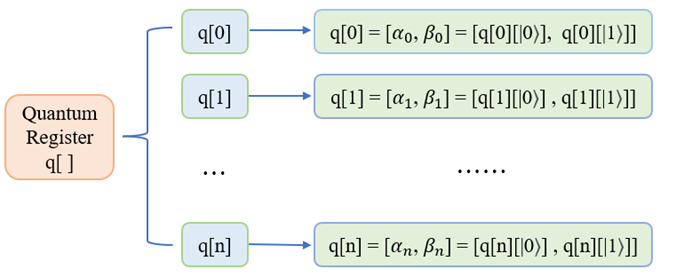}}
\caption{The probability amplitudes of a qubit in quantum register.}
\label{fig:Coefficient}
\end{figure}

\subsection{Representation of Qubit Probability Amplitudes}
In classical programming languages, an array \verb+int a[i]+ represents its elements as \verb+a[0]+, \verb+a[1]+, \ldots, \verb+a[i-1]+. In quantum programming languages like Scaffold, a quantum register is defined using the statement \texttt{qreg q[i]}, and its elements can be accessed as \verb+q[0]+, \texttt{q[1]}, \ldots, \verb+q[i-1]+. Each qubit $\ket{q}$ = $\alpha\ket{0}$ + $\beta\ket{1}$ can also be represented as [$\alpha$, $\beta$]$^\top$, where $\alpha$ and $\beta$ are the probability amplitudes of $\ket{0}$ and $\ket{1}$, respectively. To distinguish between the representation of qubit probability amplitudes and the number of qubits in a quantum register, in ScaffML, we use q[$\ket{0}$] and q[$\ket{1}$] to express the probability amplitudes $\alpha$ and $\beta$ of each qubit, respectively. Therefore, a qubit $\ket{q}$ can be expressed as $\ket{q}$ = $\alpha$$\ket{0}$ + $\beta$$\ket{1}$ = [$\alpha$, $\beta$]$^\top$ = [q[$\ket{0}$], q[$\ket{1}$]]$^\top$.

The purpose of using $\ket{0}$ and $\ket{1}$ is to indicate that a variable represents a qubit. Thus, q[$\ket{0}$] and q[$\ket{1}$] represent the probability amplitudes of a qubit $\ket{q}$. For instance, in Figure~\ref{fig:Coefficient}, the probability amplitudes $\alpha$ and $\beta$ of q[0] are represented as q[0][$\ket{0}$] and q[0][$\ket{1}$], respectively. The [0] indicates that this qubit is the first one in the quantum register q[], while [$\ket{0}$] and [$\ket{1}$] represent the probability amplitudes of $\ket{0}$ and $\ket{1}$ for q[0].

\subsection{Pre-Defined Modules in ScaffML}

To facilitate the specification of quantum gates or modules in Scaffold, we have introduced several fundamental operation modules in ScaffML, as depicted in Figure~\ref{fig:Excerpt of simple modules with ScaffML}. These modules serve as a means of specifying the pre- and postconditions of qubit states, replacing the need for multiple annotated equations. The annotated equation provided after the \texttt{//} is equivalent to the module presented above it. For instance, we utilize the predefined module \texttt{qubitselfCheck} to express the condition that the sum of the squares of the probability amplitudes should equal 1 for each qubit. The \textbf{\texttt{ensures}} keyword indicates that the module or equation defines the postconditions. Additionally, the \textbf{\texttt{old}} keyword captures the value of the variable prior to the execution of the module.

\begin{figure}[!h]
{\scriptsize
\begin{alltt}
\textbf{ensures} equal_qbit: Unchanged\{Here,Old\}(qbit, 2);
//\textbf{ensures} qbit[|0>] == \texttt{\char92}old(qbit[|0>]);
//\textbf{ensures} qbit[|1>] == \texttt{\char92}old(qbit[|1>]);

\textbf{ensures} reverse_qbit: Reverse\{Here,Old\}(qbit, 2);
//\textbf{ensures} qbit[|0>] == \texttt{\char92}old(qbit[|1>]);
//\textbf{ensures} qbit[|1>] == \texttt{\char92}old(qbit[|0>]);

\textbf{ensures} equal_qbit1: EqualRanges\{Here,Old\}(qbit1, 2, qbit2);
//\textbf{ensures} qbit1[|0>] == \texttt{\char92}old(qbit2[|0>]);
//\textbf{ensures} qbit1[|1>] == \texttt{\char92}old(qbit2[|1>]);

\textbf{ensures} Hadamard_qbit: HadamardCheck\{Here,Old\}(qbit, 2);
//\textbf{ensures} qbit[|0>] == (\texttt{\char92}old(qbit[|0>]) + 
                            \texttt{\char92}old(qbit[|1>]))*sqrt(1/2);
//\textbf{ensures} qbit[|1>] == (\texttt{\char92}old(qbit[|0>]) - 
                            \texttt{\char92}old(qbit[|1>]))*sqrt(1/2);

\textbf{ensures} Phase_qbit: PhaseCheck\{Here,Old\}(qbit, 2);
//\textbf{ensures} qbit[|0>] == \texttt{\char92}old(qbit[|0>]);
//\textbf{ensures} qbit[|1>] == \texttt{\char92}old(qbit[|1>])) * e^(i*angle);

\textbf{ensures} qbitself_qbit: qbitselfCheck(qbit);
//\textbf{ensures} pow(qbit[|0>],2) + pow(qbit[|1>],2) == 1;  

\textbf{ensures} Phase_Rx_qbit: PhaseCheck_Rx\{Here,Old\}(qbit[0], 2);
//\textbf{ensures} qbit[0][|0>] == \texttt{\char92}old(qbit[0][|0>])*cos(angle/2) - 
  \texttt{\char92}old(qbit[0][|1>])*isin(angle/2);
//\textbf{ensures} qbit[0][|1>] == \texttt{\char92}old(qbit[0][|1>])*cos(angle/2) - 
  \texttt{\char92}old(qbit[0][|0>])*isin(angle/2);

\textbf{ensures} Phase_1: PhaseCheck\{Here,Old\}(qbit[0], 2);
//\textbf{ensures} qbit[0][|0>] == \texttt{\char92}old(qbit[0][|0>])
//\textbf{ensures} qbit[0][|1>] == \texttt{\char92}old(qbit[0][|1>]) * e^(i*angle);

\end{alltt}
  \caption{Some examples of pre-defined modules in ScaffML.}
  \label{fig:Excerpt of simple modules with ScaffML}
}
\end{figure}

\subsection{Usage of Assertion}
An assertion is usually used before or after a code fragment or module to specify the desired results or behaviors of the program. ScaffML supports assertions in Scaffold programs, which can be used like assertions in ACSL. Moreover, an assertion in ScaffML satisfies the conditions of Hoare-Calculus rules~\cite{hoare1969axiomatic} as in ACSL. Note that in ScaffML, we consider that assertion is a special utilization scene of \textbf{\texttt{ensures}} in pre- and postconditions.

The basic expression of assertion is 

{\footnotesize
\begin{alltt}
              //@ \textbf{assert} P;
                Q
              //@ \textbf{assert} R;
\end{alltt}
}

\noindent
Here, \texttt{P} and \texttt{R} are logical expressions, and \texttt{Q} is a code fragment or module in Scaffold.

Figure~\ref{fig:An example of assertion} illustrates the usage of an assertion while preparing a Bell State. In Figure~\ref{fig:gate-entangle}, the first operation involved in preparing a Bell State is a Hadamard gate. We specify that the input qubit $\ket{a}$ is in the state $\frac{\ket{0}+\ket{1}}{\sqrt{2}}$ after the first operation.

\begin{figure}[!h]
{\footnotesize
\begin{alltt}
   \textbf{module} PrepareBellPair(qreg a[1], qreg b[1]) \{
      H(a[0]);
      //@ \textbf{assert} a[0][|0>] == sqrt(1/2);
      //@ \textbf{assert} a[0][|1>] == sqrt(1/2);
      CNOT(a[0],b[0]);
   \}
\end{alltt}
  \caption{An example usage of an assertion.}
  \label{fig:An example of assertion}
}
\end{figure}

\subsection{Specifying a Pauli Gate}

There are three types of Pauli gates: Pauli-X, Pauli-Y, and Pauli-Z. Let us consider the Pauli-X gate as an example to illustrate how to specify Pauli gates. The Pauli-X gate corresponds to a rotation of a single qubit around the x-axis by $\pi$ radians. Its matrix representation is:

\begin{equation}
\notag
X={\left[ \begin{array}{cc}
0 & 1\\
1 & 0
\end{array} 
\right]}
\end{equation}

\noindent
This gate transforms a qubit $\ket{q} = \alpha\ket{0} + \beta\ket{1} = [\alpha, \beta]^{\top} = [q[\ket{0}], q[\ket{1}]]^{\top}$ to $X\ket{q} = \beta\ket{0} + \alpha\ket{1} = [\beta, \alpha]^{\top} = [q[\ket{1}], q[\ket{0}]]^{\top}$.
To specify the Pauli-X gate in ScaffML, we utilize the ScaffML notation, as shown in Figure~\ref{fig:Pauli-X gate}. The \textbf{\texttt{requires valid}} clause specifies the name and size of the required array. The \textbf{\texttt{assigns}} clause indicates that the referenced locations can be modified while the array's length remains unchanged. The postcondition is then specified, ensuring that the probability amplitudes of the qubit are exchanged. Finally, we use the pre-defined module \texttt{qubitselfCheck()} to specify that the sum of the probability amplitudes' squares equals 1. This ensures the validity of the qubit state.

\begin{figure}[!h]
{\footnotesize
\begin{alltt}
/*@
  \textbf{requires valid}: \texttt{\char92}valid(input[0]+(|0>..|1>)); 

  \textbf{assigns} input[0][|0>..|1>];
  
  \textbf{ensures} reverse_input: Reverse\{Here,Old\}
                         (input[0], 2);
  \textbf{ensures} qbitself_input[0]: qbitselfCheck(input[0]);
*/
\textbf{gate} X(qreg input[1]);
\end{alltt}
  \caption{An example of specifying a Pauli-X gate.}
  \label{fig:Pauli-X gate}
}
\end{figure}

\subsection{Specifying a Hadamard Gate}

The Hadamard gate is a fundamental single-qubit operation used in quantum computing. It transforms the basis states $\ket{0}$ and $\ket{1}$ into superposition states, specifically $\frac{\ket{0}+\ket{1}}{\sqrt{2}}$ and $\frac{\ket{0}-\ket{1}}{\sqrt{2}}$, respectively. This gate creates an equal superposition of the two basis states, making it a crucial component in quantum algorithms. Mathematically, the Hadamard gate is represented by the Hadamard matrix as follows:

$$
H=\frac{1}{\sqrt{2}}{
\left[ \begin{array}{cc}
1 & 1\\
1 & -1
\end{array} 
\right ]}
$$

\noindent
Using ScaffML, we can specify the Hadamard gate as shown in Figure~\ref{fig:Hadamard Gate}. 

\begin{figure}[!h]
{\scriptsize
\begin{alltt}
/*@
  \textbf{requires valid}: \texttt{\char92}valid(input[0]+(|0>..|1>));

  \textbf{assigns} input[0][|0>..|1>];
  
  \textbf{ensures} Hadamard_input[0]: HadamardCheck\{Here,Old\}
                             (input[0], 2);
  \textbf{ensures} qbitself_input[0]: qbitselfCheck(input[0]);
*/
\textbf{gate} H(qreg t[1])
\end{alltt}
  \caption{An example of specifying a Hadamard gate.}
  \label{fig:Hadamard Gate}
}
\end{figure}

\subsection{Specifying a CNOT Gate}

The Controlled NOT (CNOT) gate is a two-qubit operation where one qubit is the control qubit, and the other is the target qubit. The CNOT gate performs the following operations: (1) It applies a Pauli-X gate to the target qubit when the control qubit is in the state $\ket{1}$, and (2) It leaves the target qubit unchanged when the control qubit is in the state $\ket{0}$. The CNOT gate can be represented by the following matrix:

$$
CNOT={
\left[ \begin{array}{cccc}
1 & 0 & 0 & 0\\
0 & 1 & 0 & 0\\
0 & 0 & 0 & 1\\
0 & 0 & 1 & 0
\end{array} 
\right ]}
$$

\noindent
Figure~\ref{fig:CNOT Gate} presents the ScaffML specification for the CNOT gate. The \textbf{\texttt{behavior}} command is used to indicate that there are multiple possible scenarios. The first scenario is when the control qubit $\ket{control[0]}$ is measured to be 0. The \textbf{measZ()} command measures the qubit to check whether it is 0 or 1. In this case, the target qubit $\ket{target[0]}$ remains unchanged. The second scenario is when the control qubit $\ket{control[0]}$ is measured to be 1. In this case, the target qubit $\ket{target[0]}$ undergoes a reversal. In other words, it passes through a Pauli-X gate.

The \textbf{\texttt{complete behaviors}} keyword indicates that, for any ranges of $\ket{control[0]}$ and $\ket{target[0]}$ that satisfy the preconditions of the contract, at least one of the specified named behaviors (in this case, \textbf{\texttt{false}} and \textbf{\texttt{true}}) applies. On the other hand, the \textbf{\texttt{disjoint behaviors}} keyword ensures that for any ranges of $\ket{control[0]}$ and $\ket{target[0]}$ that satisfy the preconditions of the contract, at most one of the specified named behaviors applies. By utilizing \textbf{\texttt{complete behaviors}} and \textbf{\texttt{disjoint behaviors}}, we ensure that the postcondition of the module accurately corresponds to one and only one scenario.

\begin{figure}[h]
{\scriptsize
\begin{alltt}
/*@
  \textbf{requires valid}: \texttt{\char92}valid(control[0]+(|0>..|1>));
  \textbf{requires valid}: \texttt{\char92}valid(target[0]+(|0>..|1>));

  \textbf{assigns} control[0][|0>..|1>];
  \textbf{assigns} target[0][|0>..|1>];

  \textbf{behavior false}:
  \textbf{assumes} measZ(control[0]) == 0;
  \textbf{ensures} equal_control[0]: Unchanged\{Here,Old\}(control[0],2);
  \textbf{ensures} equal_target[0]: Unchanged\{Here,Old\}(target[0],2);

  \textbf{behavior true}:
  \textbf{assumes} measZ(control[0]) == 1;
  \textbf{ensures} equal_control[0]: Unchanged\{Here,Old\}(control[0],2);
  \textbf{ensures} reverse_target[0]: Reverse\{Here,Old\}(target[0],2);

  \textbf{complete behaviors};
  \textbf{disjoint behaviors};

  \textbf{ensures} qbitself_control[0]: qbitselfCheck(control[0]);
  \textbf{ensures} qbitself_target[0]: qbitselfCheck(target[0]);
*/
\textbf{gate} CNOT(qreg target[1], qbit control[1]) 
\end{alltt}
  \caption{An example of specifying a CNOT gate.}
  \label{fig:CNOT Gate}
}
\end{figure}

\subsection{Specifying a Rotation Gate}

The rotation operators are generated by exponentiating the Pauli matrices according to the formula $exp(iAx) = \cos(x)I + i\sin(x)A$, where A represents one of the three Pauli matrices ($R_x$, $R_y$, and $R_z$). Specifically, the $R_x$ gate is a single-qubit rotation operation around the x-axis by an angle $\theta$ (in radians). The $R_x$ gate can be represented by the $R_x$ matrix:
\begin{equation}
\notag
R_x(\theta)=exp(-iX\theta/2)={
\left[ \begin{array}{cc}
cos(\frac{\theta}{2}) & -isin(\frac{\theta}{2})\\
-isin(\frac{\theta}{2}) & cos(\frac{\theta}{2})
\end{array} 
\right ]}
\end{equation}

\noindent
In ScaffML, we can specify the $R_x$ gate as shown in Figure~\ref{fig:$R_x$ Gate}. The specification ensures that the output of the gate matches the behavior of the $R_x$ matrix operation.

\begin{figure}[h]
{\scriptsize
\begin{alltt}
/*@
  \textbf{requires valid}: \texttt{\char92}valid(input[0]+(|0>..|1>));

  \textbf{assigns} input[0][|0>..|1>];
  
  \textbf{ensures} Phase_Rx: PhaseCheck_Rx_input\{Here,Old\}
                    (input[0], 2);
  \textbf{ensures} qbitself_input[0]: qbitselfCheck(input[0]);
*/
\textbf{gate} Rx(qreg input[1], float angle)
\end{alltt}
  \caption{An example of specifying a $R_x$ gate.}
  \label{fig:$R_x$ Gate}
}
\end{figure}

\subsection{Specifying a Phase Shift Gate}

The phase shift gate is a family of single-qubit gates that modify the phase of the quantum state. It maps the basis states $\ket{0}\mapsto\ket{0}$ and $\ket{1}\mapsto e^{i\varphi} \ket{1}$, where $\varphi$ is the phase shift angle. The probability amplitudes of $\ket{0}$ and $\ket{1}$ remain unchanged after applying the phase shift gate, but the phase of the quantum state is modified. The phase shift gate can be represented by the phase shift matrix as follows:

\begin{equation}
\notag
P(\varphi)={
\left[ \begin{array}{cc}
1 & 0\\
0 & e^{i\varphi}
\end{array} 
\right ]}
\end{equation}

\noindent
Using ScaffML, we can specify the phase shift gate as shown in Figure~\ref{fig:Phase shift Gate}. The coefficient of $\ket{0}$ keeps itself and the coefficient of $\ket{1}$ rotated angle $\varphi$, so it multiplied $e^{i\varphi}$.

\begin{figure}[h]
{\scriptsize
\begin{alltt}
/*@
  \textbf{requires valid}: \texttt{\char92}valid(input[0]+(|0>..|1>));

  \textbf{assigns} input[0][|0>..|1>];
  
  \textbf{ensures} Phase_input[0]: PhaseCheck\{Here,Old\}
                          (input[0], 2);
  \textbf{ensures} qbitself_input[0]: qbitselfCheck(input[0]);
*/
\textbf{gate} Phase(qreg input[1], float angle)
\end{alltt}
  \caption{An example of specifying a Phase shift gate.}
  \label{fig:Phase shift Gate}
}
\end{figure}

\subsection{Specifying a SWAP Gate}

The SWAP gate is a two-qubit operation that swaps the states of the two qubits involved. In terms of basis states, the SWAP gate can be represented as:

\begin{equation}
\notag
\textit{SWAP}={
\left[ \begin{array}{cccc}
1 & 0 & 0 & 0\\
0 & 0 & 1 & 0\\
0 & 1 & 0 & 0\\
0 & 0 & 0 & 1
\end{array} 
\right ]}
\end{equation}

\noindent
In ScaffML, we specify the SWAP gate as shown in Figure~\ref{fig:SWAP Gate}. The postcondition states that the probability amplitudes of $\ket{input1}$ and $\ket{input2}$ are swapped. The command \texttt{EqualRanges} ensures that the elements of the first array are equal to the elements of the second array, with \texttt{Here} and \texttt{old} used to represent the values of the array after and before the SWAP gate, respectively. The number \texttt{2} indicates that each array has two elements.

\begin{figure}[!h]
{\scriptsize
\begin{alltt}
/*@
  \textbf{requires valid}: \texttt{\char92}valid(input1[0]+(|0>..|1>));
  \textbf{requires valid}: \texttt{\char92}valid(input2[0]+(|0>..|1>));

  \textbf{assigns} input1[0][|0>..|1>];
  \textbf{assigns} input2[0][|0>..|1>];

  \textbf{ensures} equal_input1[0]: EqualRanges\{Here,Old\}
                           (input1[0], 2, input2[0]);
  \textbf{ensures} equal_input2[0]: EqualRanges\{Old,Here\}
                           (input2[0], 2, input2[0]);
  \textbf{ensures} qbitself_input2[0]: qbitselfCheck(input1[0]);
  \textbf{ensures} qbitself_input1[0]: qbitselfCheck(input2[0]);
*/
\textbf{gate} SWAP(qreg input1[1], qreg input2[1]) 
\end{alltt}
  \caption{An example of specifying a SWAP gate.}
  \label{fig:SWAP Gate}
}
\end{figure}

\section{Specifying Programs in ScaffML}\label{sec:ScaffML}

In Scaffold, modules are fundamental implementation units, similar to C functions. ScaffML allows the specification of individual module properties in a Scaffold program using module specifications, and the specifications of all modules in the program form a specification of the entire program.

The module specification in ScaffML resembles a function specification in ACSL or a method specification in JML. It includes two formulas: a precondition and a postcondition (declared using the \texttt{requires} and \texttt{ensures} clauses, respectively). These components together constitute the module's specification, enabling the verification of the module's code.

In this section, we demonstrate using ScaffML to specify two commonly used modules in quantum algorithms: the Bell state and the Quantum Fourier Transform (QFT).

\subsection{Specifying the Bell State module}

\begin{figure}[th]
\centerline{\includegraphics[width=0.6\linewidth]{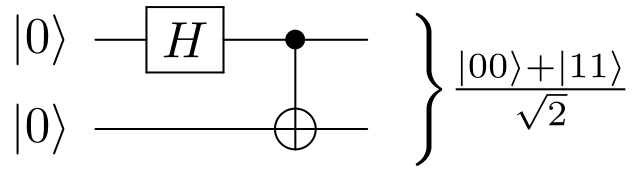}}
\caption{Quantum circuit for the Bell state.}
\label{fig:gate-entangle}
\end{figure}

The Bell state is a fundamental entangled quantum state. It represents the superposition of two basis states, $\frac{\ket{00}+\ket{11}}{\sqrt{2}}$, as depicted in Figure~\ref{fig:gate-entangle}. When one qubit is measured, its value becomes immediately correlated with the value of the other qubit. In other words, both qubits will have the same measurement outcome, either 0 or 1.

To specify the behavior of the Bell state, Figure~\ref{fig:Specifying the Bell State} provides its specification. Initially, we ensure that the input qubits $\ket{a}$ and $\ket{b}$ are in the state of 1$\ket{0}$ + 0$\ket{1}$. After creating the Bell pair, two possible scenarios arise. In the first scenario, the Bell pair is in the state $\ket{00}$, where measuring $\ket{a}$ and $\ket{b}$ yields 0, indicating that the CNOT gate has not operated. In the second scenario, the Bell pair is in the state $\ket{11}$, where measuring $\ket{a}$ and $\ket{b}$ results in 1, indicating that the CNOT gate has successfully applied its transformation.

\begin{figure}[!h]
{\scriptsize
\begin{alltt}
      /*@
        \textbf{requires valid}: \texttt{\char92}valid(a[0]+(|0>..|1>));
        \textbf{requires valid}: \texttt{\char92}valid(b[0]+(|0>..|1>));

        \textbf{assigns} a[0][|0>..|1>];
        \textbf{assigns} b[0][|0>..|1>];
  
        \textbf{ensures} \texttt{\char92}old(a[0][|0>]) == 1;
        \textbf{ensures} \texttt{\char92}old(a[0][|1>]) == 0;
        \textbf{ensures} \texttt{\char92}old(b[0][|0>]) == 1;
        \textbf{ensures} \texttt{\char92}old(b[0][|1>]) == 0;
        \textbf{ensures} a[0][|0>] == sqrt(1/2);
        \textbf{ensures} a[0][|1>] == sqrt(1/2);
        \textbf{ensures} b[0][|0>] == sqrt(1/2);
        \textbf{ensures} b[0][|1>] == sqrt(1/2);

        \textbf{behavior CNOTfalse}:
            \textbf{assumes} measZ(a[0]) == 0;
            \textbf{ensures} measZ(b[0]) == 0;

        \textbf{behavior CNOTtrue}:
            \textbf{assumes} measZ(a[0]) == 1;
            \textbf{ensures} measZ(b[0]) == 1;

        \textbf{complete behaviors};
        \textbf{disjoint behaviors};

        \textbf{ensures} qbitself_a[0]: qbitselfCheck(a[0]);
        \textbf{ensures} qbitself_b[0]: qbitselfCheck(b[0]);
      */
      \textbf{module} PrepareBellPair(qreg a[1], qreg b[1]) \{
            H(a[0]);
            CNOT(a[0],b[0]);
      \}
\end{alltt}
  \caption{Specifying the Bell state module.}
  \label{fig:Specifying the Bell State}
}
\vspace*{-5mm}
\end{figure}

\begin{figure*}[th]
\centerline{\includegraphics[width=0.85\linewidth]{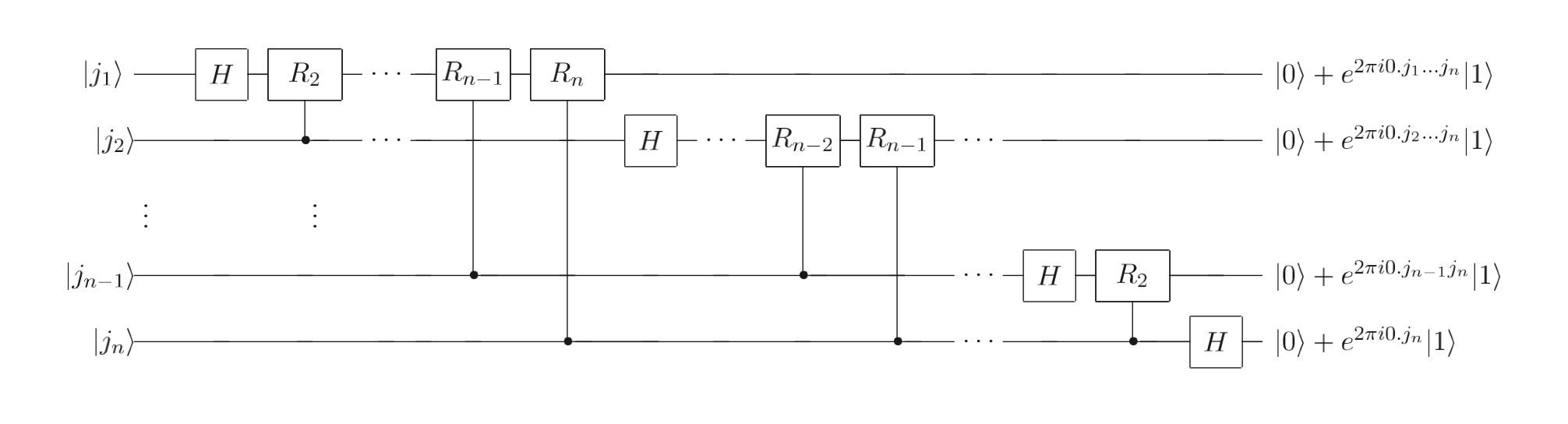}}
\caption{Quantum circuit for the Quantum Fourier Transform (QFT) algorithm.}
\label{fig:QFT_circuit}
\end{figure*}

\subsection{Specifying the QFT Modules}

The Quantum Fourier Transform (QFT) is a powerful mathematical operation in quantum computing that performs a discrete Fourier transform on a quantum state. It decomposes the original equation into a simpler product of multiple unitary matrices. The quantum circuit representation of the QFT, shown in Figure~\ref{fig:QFT_circuit}, consists of several Hadamard gates and controlled phase-shifting gates.

In Figure~\ref{fig:Specifying the QFT}, we provide the specification of the QFT program. The program consists of two modules: \texttt{controlledRd} and \texttt{QFT}. These modules can be specified separately using ScaffML. The module \texttt{QFTcheck} in Figure~\ref{fig:Specifying the QFT} demonstrates how we can define a module to be specified similarly to pre-defined modules. Using the defined module \texttt{QFTcheck}, we can specify the behavior of the \texttt{QFT} module. Within \texttt{QFTcheck}, the \texttt{width} parameter represents the number of qubits, and \texttt{M\_PI} represents the total rotation angle of the Hilbert space formed by the input qubits. By analyzing the periodicity of the input Hilbert space, we can determine the states of each qubit.

The postcondition of the \texttt{QFT} module ensures that the output Hilbert space exhibits the periodicity of the input Hilbert space. We analyze the output of the \texttt{QFT} module and specify the states of each qubit. The output Hilbert space consists of qubits that are in the state of 1$\ket{0}$ + 0$\ket{1}$ and qubits that are in the state of 0$\ket{0}$ + 1$\ket{1}$. However, we only specify the periodic inputs, as irregular inputs are meaningless for the QFT operation.

\begin{figure}[h]
{\scriptsize
\begin{alltt}
/*@
  \textbf{requires valid}: \texttt{\char92}valid(control[0]+(|0>..|1>));
  \textbf{requires valid}: \texttt{\char92}valid(target[0]+(|0>..|1>));
  \textbf{assigns} control[0][|0>..|1>];
  \textbf{assigns} target[0][|0>..|1>];
  \textbf{behavior false}:
  \textbf{assumes} measZ(control[0]) == 0;
  \textbf{ensures} equal_control[0]: Unchanged\{Here,Old\}(control[0],2);
  \textbf{ensures} equal_target[0]: Unchanged\{Here,Old\}(target[0],2);
  \textbf{behavior true}:
  \textbf{assumes} measZ(control[0]) == 1;
  \textbf{ensures} equal_control[0]: Unchanged\{Here,Old\}(control[0],2);
  \textbf{ensures} target[0][|0>] == 
                \texttt{\char92}old(target[0][|0>])*cos(PI/pow(2,d+1))- 
                \texttt{\char92}old(target[0][|1>])*isin(PI/pow(2,d+1));
  \textbf{ensures} target[0][|1>] == 
                \texttt{\char92}old(target[0][|1>])*cos(PI/pow(2,d+1))+ 
                \texttt{\char92}old(target[0][|0>])*isin(PI/pow(2,d+1));
  \textbf{complete behaviors};
  \textbf{disjoint behaviors};
  \textbf{ensures} qbitself_control[0]: qbitselfCheck(control[0]);
  \textbf{ensures} qbitself_target[0]: qbitselfCheck(target[0]);
*/
\textbf{module} controlledRd(qreg target[1], qreg control[1], int d)\{
  float angle;
  angle = PI/pow(2,d);   
  //standard rotation gate from the library
  controlledRz(target[0], control[0], angle);
\}

/*@
  \textbf{requires valid}: \texttt{\char92}valid(qbits[]+(|0>..|1>));
  \textbf{assigns} qbits[][|0>..|1>];                
  \textbf{module} QFTCheck(qbits[], width, M_PI) \{
  int r = M_PI/pi;
  for ( int s = width-1; s >= 0; s-- ) \{
  if ( power(2,s) <= M_PI/pi )
  \{
  //\textbf{ensures} qbit[s][|1>] == 1;
  //\textbf{ensures} qbit[s][|0>] == 0;
  r = r - power(2,s);
  \}
  else
  \{
  //\textbf{ensures} qbit[s][|0>] == 1;
  //\textbf{ensures} qbit[s][|1>] == 0;
  \}
  \}
  \}
  \textbf{ensures} QFTCheck_qbits[]: QFTCheck(qbits[], width, M_PI);
  \textbf{ensures} qbitself_qbits[]: qbitselfCheck(qbits[]);
*/
\textbf{module} QFT(qreg qbits[width]) \{
  // Termination condition
  if(length(qbits) == 1) \{
  return;
  \}    
  // Recursively call
  QFT(qbits[0..length(qbits) - 2]);    
  // Hadamard
  H(qbits[length(qbits)-1]);    
  // Rotation chain
  int i=0;  
  for(i = 0; i < length(data)-1; i++) \{
    controlledRd(qbits[i], data[length(qbits)-1], 
                    length(qbits)-i-1);
  \}
\}
\end{alltt}
\caption{Specifying the QFT modules.}
  \label{fig:Specifying the QFT}
}
\end{figure}

\begin{figure}[h]
\centerline{\includegraphics[width=0.9\linewidth]{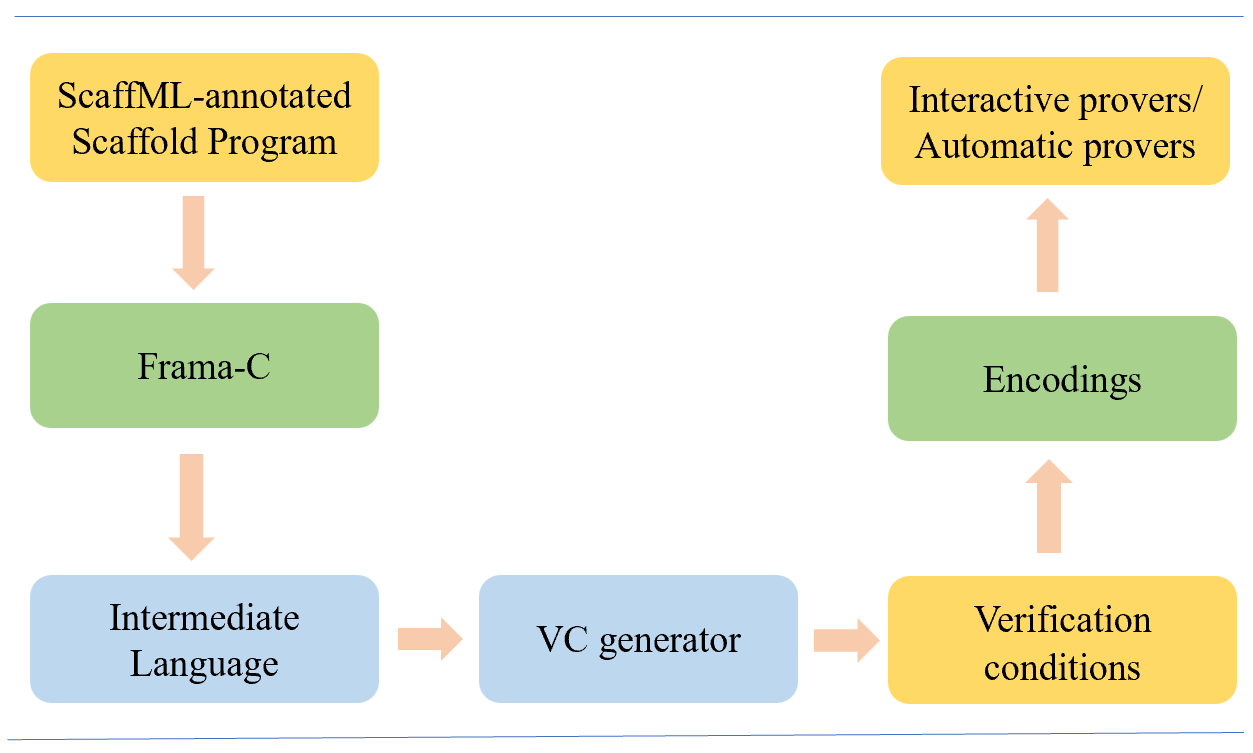}}
\caption{The Workflow of ScaffML.}
\label{fig:Workflow}
\end{figure}

\section{Tool Support for ScaffML working system}
\label{sec:tool-support}

Figure~\ref{fig:Workflow} depicts the workflow of the ScaffML working system. Initially, a Scaffold program, along with its ScaffML specification, undergoes translation into an intermediate code written in a standardized intermediate language. Subsequently, the code is passed to a verification condition (VC) generator responsible for generating the necessary verification conditions. Finally, these verification conditions are received by either an interactive prover or an automatic prover, such as Coq or CVC3, for further processing.

The decision to base ScaffML on ACSL stems from the aim of leveraging existing tools developed for ACSL. To accomplish this, we are developing a dedicated tool capable of automatically transforming Scaffold programs with their corresponding ScaffML specifications into a universally accepted intermediate language, as illustrated in Figure~\ref{fig:Workflow}. Our tool builds upon a modified version of Frama-C~\cite{Frama-C}, enabling the deductive verification of Scaffold programs. Prior to translation into the embedded intermediate language, Scaffold files annotated with ScaffML undergo syntax error checks performed by the modified Frama-C core. The verification condition generator then produces the verification conditions, which are subsequently processed by an interactive or automatic verifier, such as Coq or CVC3.

\section{Related Work}\label{sec:related-work}

There has been significant progress in the field of generic specification languages and BISLs~\cite{hatcliff2012behavioral}. Some widely used generic specification languages include Z~\cite{spivey1992z}, VDM~\cite{jones1990systematic}, B~\cite{abrial2005b}, and Larch~\cite{guttag1993larch}. Several BISLs based on Larch have been developed, each tailored to specific programming languages. For instance, LCL (for C)\cite{guttag1993larch} and Larch/C++\cite{leavens1996overview} provide specification support for their respective languages. In addition to the Larch family, Meyer has done notable work on the programming language Eiffel, which has significantly contributed to applying formal methods to object-oriented programs~\cite{meyer2010object}. In Eiffel, unlike Larch-style interface specification languages, Boolean expressions can be utilized to specify pre- and postconditions for operations on Abstract Data Types (ADTs) written in Eiffel. This means that program expressions can be employed within pre- and postconditions. Furthermore, class invariants can be used in Eiffel to specify the global properties of class instances.

As Java has gained popularity as an object-oriented programming language in recent years, several BISLs have been specifically designed for Java. Examples of these include JML~\cite{leavens2006preliminary} and AAL~\cite{khurshid2002analyzable}. JML allows the specification of assertions for Java classes and interfaces, providing extensive expressive power for Java module specification. On the other hand, AAL is an annotation language tailored for annotating and checking Java programs. Similar to JML, AAL supports runtime assertion checking but offers comprehensive static checking capabilities for Java programs. AAL translates annotated Java programs into Alloy~\cite{jackson2002alloy}, a first-order logic with relational operators, and leverages Alloy's SAT solver-based automatic analysis technique to verify Java programs.

Although the generic specification languages and BISLs mentioned above can specify programs written in various classical programming languages, they are not designed to specify programs written in quantum programming languages such as Scaffold~\cite{Beyond}.

Singhal et al.~\cite{Beyond} proposed an outline of a specification language to specify quantum programs to support modular reasoning about these programs. The basic idea is to introduce Ghost variables into the specification to represent any missing parts of an entangled system, thus making modular reasoning about entanglement possible. Both their work and ours support modular reasoning about quantum programs, but the difference is that they focus on a generic specification language, whereas we focus on a BISL tailored to a specific quantum programming language (i.e., Scaffold).

Ying~\cite{ying2012floyd} proposed the quantum Floyd-Hoare logic to support the specification and verification of quantum programs. The goal is to prove the completeness and correctness of the logic of quantum programs by using the weakest precondition. Ying's quantum Floyd-Hoare logic is dedicated to providing a general method for specifying and verifying quantum programs, while our ScaffML takes a more specific approach by designing a specification language tailored to the Scaffold quantum programming language to support lightweight specification and verification of Scaffold programs.

\section{Conclusion and Future Work}
\label{sec:conclusion}
In this paper, we have presented ScaffML, a behavior interface specification language tailored to Scaffold, and discussed the goals of ScaffML and the overall specification approach. ScaffML is an extension to ACSL, a BISL for C, for specifying Scaffold programs. ScaffML uses the same way as ACSL to specify Scaffold classical modules and introduces several new notations to specify Scaffold quantum modules with pre- and postconditions. Unlike classical specification languages, ScaffML supports specifying quantum modules that contain quantum superposition and entanglement.

On one hand, ScaffML provided a way to specify Scaffold programs with assertions (pre- and postconditions and module invariants), supporting runtime checking such as debugging and testing of Scaffold programs. On the other hand, ScaffML offered the possibility of fully automatic compile-time analysis for Scaffold programs, such as checking the code of a Scaffold module against its specification. Additionally, examples of ScaffML were provided to demonstrate how to specify a set of quantum gates in Scaffold, the entanglement state (Bell state), and the QFT quantum algorithm.

The work presented in this paper is preliminary, and further efforts are needed to make ScaffML a practical tool. Our future work will focus on the following aspects:

\begin{itemize}
\item Refining the proposed specification constructs for Scaffold, ensuring their effectiveness and usability.

\item Developing a transformation tool that automatically converts Scaffold programs along with their ScaffML specifications into verification conditions (VC) using VC generators. These VCs can then be processed by interactive or automatic provers such as Coq~\cite{Formalproof} or CVC3~\cite{barrett2007cvc3} to verify Scaffold programs formally.

\item Establishing formal semantics for ScaffML to enable static analysis, checking, and testing of Scaffold programs.

\item Exploring the design of BISLs for other quantum programming languages to extend the applicability of BISLs to a broader range of quantum programming languages.

\item Conducting case studies using ScaffML to specify complex Scaffold programs, particularly for implementing renowned quantum algorithms such as Shor's and Grover's algorithms.

\end{itemize}


\bibliographystyle{plain}
\bibliography{qse-bibliography}

\begin{thebibliography}{10}

\bibitem{Frama-C}
Frama-{C} - framework for modular analysis of {C} programs.
\newblock \url{https://frama-c.com/}.

\bibitem{abhari2012scaffold}
Ali~J Abhari, Arvin Faruque, Mohammad~J Dousti, Lukas Svec, Oana Catu, Amlan
  Chakrabati, Chen-Fu Chiang, Seth Vanderwilt, John Black, and Fred Chong.
\newblock Scaffold: Quantum programming language.
\newblock Technical report, Department of Computer Science, Princeton
  University, 2012.

\bibitem{abrial2005b}
Jean-Raymond Abrial and Jean-Raymond Abrial.
\newblock {\em The B-book: assigning programs to meanings}.
\newblock Cambridge university press, 2005.

\bibitem{aleksandrowicz2019qiskit}
Gadi Aleksandrowicz, Thomas Alexander, Panagiotis Barkoutsos, Luciano Bello,
  Yael Ben-Haim, David Bucher, F~Jose Cabrera-Hern{\'a}ndez, Jorge
  Carballo-Franquis, Adrian Chen, Chun-Fu Chen, et~al.
\newblock Qiskit: An open-source framework for quantum computing.
\newblock {\em Accessed on: Mar}, 16, 2019.

\bibitem{barrett2007cvc3}
Clark Barrett and Cesare Tinelli.
\newblock Cvc3.
\newblock In {\em International Conference on Computer Aided Verification},
  pages 298--302. Springer, 2007.

\bibitem{baudin2008acsl}
Patrick Baudin, Jean-Christophe Filli{\^a}tre, Claude March{\'e}, Benjamin
  Monate, Yannick Moy, and Virgile Prevosto.
\newblock {ACSL}: {ANSI/ISO} {C} specification.
\newblock 2008.

\bibitem{Formalproof}
Georges Gonthier.
\newblock Formal proof—the four-color theorem.
\newblock {\em Notices of the American Mathematical Society},
  55(11):1382--1393, 2008.

\bibitem{green2013quipper}
Alexander~S Green, Peter~LeFanu Lumsdaine, Neil~J Ross, Peter Selinger, and
  Beno{\^\i}t Valiron.
\newblock Quipper: a scalable quantum programming language.
\newblock In {\em Proceedings of the 34th ACM SIGPLAN conference on Programming
  language design and implementation}, pages 333--342, 2013.

\bibitem{guttag1993larch}
John~V Guttag and James~J Horning.
\newblock {\em Larch: languages and tools for formal specification}.
\newblock Springer Science \& Business Media, 1993.

\bibitem{hatcliff2012behavioral}
John Hatcliff, Gary~T Leavens, K~Rustan~M Leino, Peter M{\"u}ller, and Matthew
  Parkinson.
\newblock Behavioral interface specification languages.
\newblock {\em ACM Computing Surveys (CSUR)}, 44(3):1--58, 2012.

\bibitem{hoare1969axiomatic}
Charles Antony~Richard Hoare.
\newblock An axiomatic basis for computer programming.
\newblock {\em Communications of the ACM}, 12(10):576--580, 1969.

\bibitem{hoare1978proof}
Charles Antony~Richard Hoare.
\newblock Proof of correctness of data representations.
\newblock In {\em Programming methodology}, pages 269--281. Springer, 1978.

\bibitem{jackson2002alloy}
Daniel Jackson.
\newblock Alloy: a lightweight object modelling notation.
\newblock {\em ACM Transactions on Software Engineering and Methodology
  (TOSEM)}, 11(2):256--290, 2002.

\bibitem{jones1990systematic}
Cliff~B Jones.
\newblock {\em Systematic software development using VDM}, volume~2.
\newblock Englewood Cliffs: Prentice Hall, 1990.

\bibitem{kernighan1988c}
Brian~W Kernighan and Dennis~M Ritchie.
\newblock {\em The C programming language}.
\newblock Prentice Hall, 1988.

\bibitem{khurshid2002analyzable}
Sarfraz Khurshid, Darko Marinov, and Daniel Jackson.
\newblock An analyzable annotation language.
\newblock In {\em Proceedings of the 17th ACM SIGPLAN conference on
  Object-oriented programming, systems, languages, and applications}, pages
  231--245, 2002.

\bibitem{leavens1996overview}
Gary~T Leavens.
\newblock An overview of {L}arch/{C}++: Behavioral specifications for {C}++
  modules.
\newblock {\em Object-Oriented Behavioral Specifications}, pages 121--142,
  1996.

\bibitem{leavens2006preliminary}
Gary~T Leavens, Albert~L Baker, and Clyde Ruby.
\newblock Preliminary design of {JML}: A behavioral interface specification
  language for {J}ava.
\newblock {\em ACM SIGSOFT Software Engineering Notes}, 31(3):1--38, 2006.

\bibitem{meyer2010object}
Bertrand Meyer.
\newblock {\em Object-oriented software construction}.
\newblock Interactive Software Engineering (ISE) Inc., 2010.

\bibitem{Beyond}
K~Singhal, R~Rand, and M~Amy.
\newblock Beyond separation: Toward a specification language for modular
  reasoning about quantum programs.
\newblock {\em Programming Languages for Quantum Computing (PLanQC) 2022 Poster
  Abstract}, 2022.

\bibitem{spivey1992z}
J~Michael Spivey and JR~Abrial.
\newblock {\em The Z notation}.
\newblock Hemel Hempstead: Prentice Hall, 1992.

\bibitem{svore2018q}
Krysta Svore, Alan Geller, Matthias Troyer, John Azariah, Christopher Granade,
  Bettina Heim, Vadym Kliuchnikov, Mariia Mykhailova, Andres Paz, and Martin
  Roetteler.
\newblock Q\#: Enabling scalable quantum computing and development with a
  high-level dsl.
\newblock In {\em Proceedings of the Real World Domain Specific Languages
  Workshop 2018}, pages 1--10, 2018.

\bibitem{ying2012floyd}
Mingsheng Ying.
\newblock Floyd--hoare logic for quantum programs.
\newblock {\em ACM Transactions on Programming Languages and Systems (TOPLAS)},
  33(6):1--49, 2012.

\end{thebibliography}


\vspace*{-1mm}
\appendix

\section{Scaffold}

Scaffold~\cite{abhari2012scaffold} is a quantum programming language developed by a team from Princeton University and other institutions. It enables the programming of computational operations and data structures involved in quantum algorithms, which can then be compiled into a machine-executable form. The team also developed ScaffCC, the compiler for Scaffold, which can compile Scaffold source code into instructions that can be executed on QX, a quantum simulator developed by QuTech Labs.

Scaffold extends the C language, introducing new data types such as \verb+qbit+ and \verb+cbit+, and defining quantum operations, including Pauli-X gates, Hadamard gates, and other quantum logic gates. A Scaffold program consists of both a classical part and a quantum part. The classical part includes classical data types and control structures, while the quantum part incorporates quantum data types and operations.

A complete Scaffold program typically comprises multiple modules. Since quantum circuits are always "reversible," these modules must satisfy certain requirements to be executed on a quantum device. They must either consist solely of unitary quantum operations or be compilable into unitary quantum operation instructions.

To compile classical modules into unitary quantum instructions, Scaffold includes the CTQG module, which can compile classical circuits into an instruction set consisting of NOT gates, controlled-NOT (CNOT) gates, and Toffoli gates. {For example, when computing the addition of \texttt{a+b}, if \verb+N+-bit binary numbers can represent both \verb+a+ and \verb+b+, the classical instruction can be composed of \verb+6N-3+ CNOT gates and \verb+2N-2+ Toffoli gates without the need for auxiliary qubits. Consequently, user-defined operations such as classical addition in Scaffold are ultimately compiled and executed as a set of quantum operation instructions.

In the following sections, we briefly introduce the syntactic details Scaffold adds to the classical programming language.

\subsection{Quantum Data Types}
The fundamental quantum data type in Scaffold is the quantum register, represented by the \textbf{\texttt{qreg}} keyword. A quantum register can be declared using the statement \textbf{\texttt{qreg qs[n]}}, where \texttt{n} specifies the number of qubits in the register. Even if \texttt{n=1}, it is still considered a quantum register rather than an individual qubit. Alternatively, multiple quantum registers can be declared as part of a single quantum structure using the \textbf{\texttt{qstruct}} keyword:

{\footnotesize
\begin{alltt}
              \textbf{qstruct} struct1 \{
                \textbf{qreg} first[10];
                \textbf{qreg} second[10];
              \};
\end{alltt}
}

\noindent
In this case, the quantum structure \textbf{\texttt{struct1}} contains two quantum registers: \texttt{first} and \texttt{second}. To access the first qubit of the \texttt{second} register within the \textbf{\texttt{struct1}} structure, the following statements can be used:

{\footnotesize
\begin{alltt}
                \textbf{struct1} qst;
                qst.second[0];
\end{alltt}
}

\begin{figure}[t]
\center{
\centerline{\includegraphics[width=1.2\linewidth]{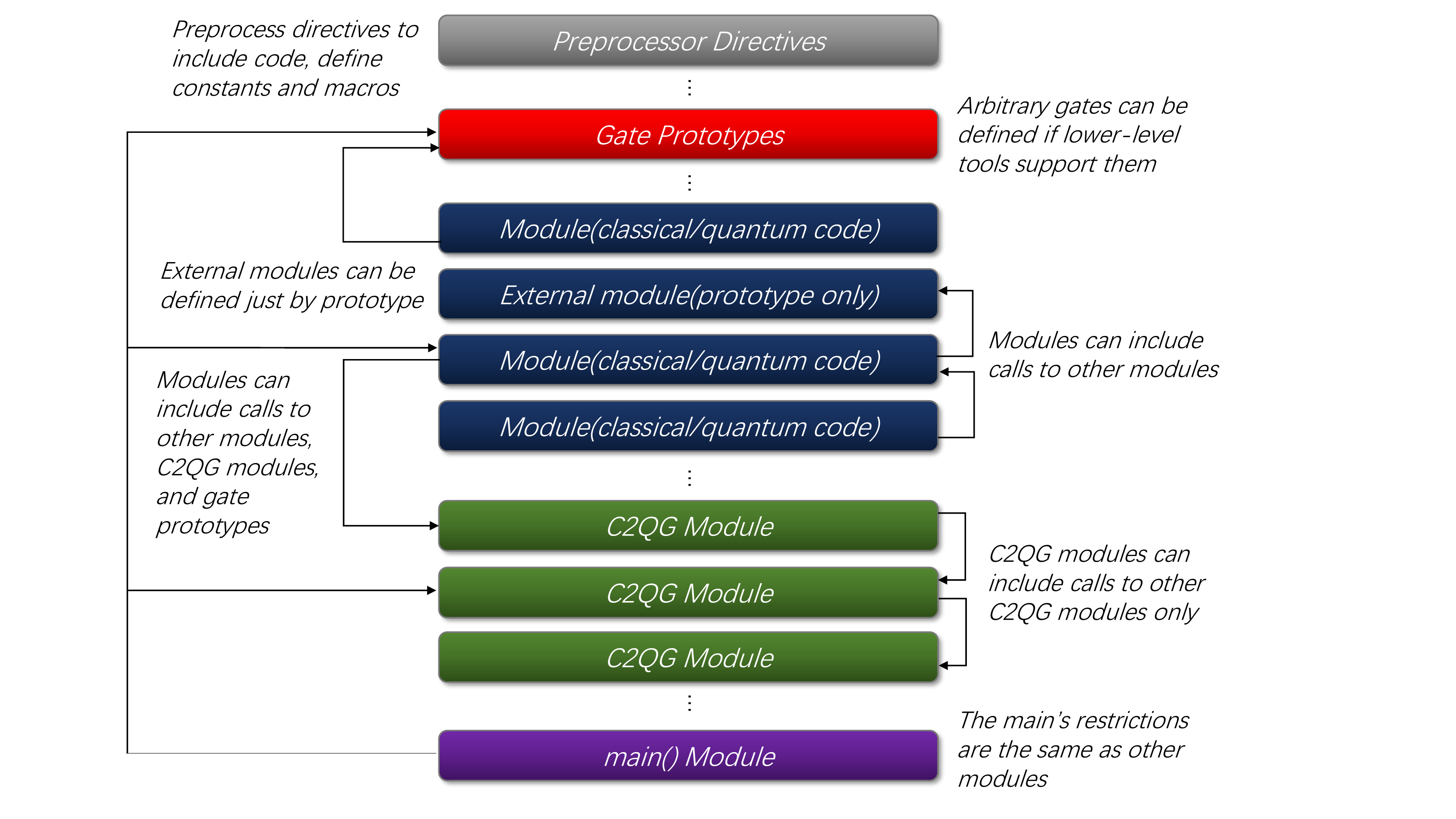}}
\caption{The whole structure of a Scaffold program~\cite{abhari2012scaffold}.}
\label{fig:scaffold-structure}
}
\end{figure}

\subsection{Quantum Gates}
The quantum gate operations in Scaffold can be categorized into two types based on their implementation: built-in quantum gates in the standard library and quantum gates defined through gate prototype functions. The first type of gate function only requires the inclusion of the \verb+gates.h+ header file. The second type of gate operation requires its definition through a gate prototype function. The syntax for defining such a gate operation is as follows:

{\scriptsize
\begin{alltt}

 \textbf{gate} gatename(\textbf{type_1} parameter_1, ..., \textbf{type_n} parameter_n);

\end{alltt}
}

\noindent
The above statement defines a gate operation named \verb+gatename+ that accepts \verb+n+ arguments. The data types of these arguments can be quantum registers or classical unsigned integers, characters, floating-point numbers, and double-precision floating-point numbers. These arguments must be passed to the function by reference (except for classical data types with the \verb+const+ keyword, which can be passed by value). Once the gate operation \verb+gatename+ is defined in a module, it can be invoked using the following statement:

{\footnotesize
\begin{alltt}
    gatename(parameter_1, ..., parameter_n);
\end{alltt}
}

\begin{figure}[h]
{\scriptsize
\begin{alltt}
    //module prototypes. They are defined elsewhere
    \textbf{module} U (qreg input[4], int n);
    \textbf{module} V (qreg input[4]);
    \textbf{module} W (qreg input[4], float p);
        
    //Quantum control primitive
    \textbf{module} control_example(qreg input[4]) \{
     \textbf{if} (control_1[0]==1 && control_2[0]==1) 
        \{ U(input); \}
     \textbf{else if} (control_1{0}==1 && control_2[0]==0) 
        \{ V(input); \}
     \textbf{else} 
        \{ W(input); \}
    \}
\end{alltt}
  \caption{Example quantum control primitive for controlled execution of 3 different modules, U, V, and W~\cite{abhari2012scaffold}}
  \label{fig:control-program}
}
\end{figure}

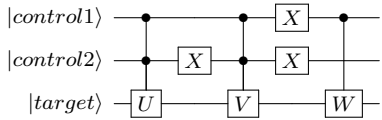
\begin{figure}[h] 
\footnotesize{
\centerline{
\Qcircuit @C=0.8em @R=0.75em {
   \lstick{\ket{control1}}   &  \ctrl{+2}   &   \qw       & \ctrl{+2}  &   \gate{X} & \ctrl{+2} &  \qw    \\
   \lstick{\ket{control2}}   &  \ctrl{+1}   &   \gate{X}  & \ctrl{+1}  &   \gate{X} & \qw       &  \qw    \\
   \lstick{\ket{target}}      &  \gate{U}    &   \qw       & \gate{V}   &   \qw      & \gate{W}  &  \qw    \\
}
}
\caption{The quantum circuit corresponding to the program in Figure~\ref{fig:control-program}.}\label{fig:quantum-circuit}
}
\end{figure}

\subsection{Loops and Control Structures}
Similar to the C language, Scaffold supports \verb+if+, \verb+switch+, and \verb+loop+ statements, but their control conditions can only involve classical information. Scaffold also introduces a quantum control statement, where the control condition includes qubits. During the evaluation of the quantum control condition, the quantum state of the qubit is determined, and the code is executed accordingly. For instance, Figure~\ref{fig:control-program} presents a simple Scaffold program excerpt from~\cite{abhari2012scaffold}, illustrating the quantum control primitive for the controlled execution of three different modules.
This program determines which code to execute based on the quantum states of the control qubits \verb+control_1[0]+ and \verb+control_2[0]+. Following the principle of delayed measurement, the code is compiled into the circuit depicted in Figure~\ref{fig:quantum-circuit}. It is important to note that the qubits used in the control condition cannot be reused in the program.

\subsection{Modules}
Scaffold supports the modular design, which enhances readability, maintainability, and other essential features. A complete program in Scaffold typically comprises one or more modules, each responsible for a specific task and capable of exchanging data of classical or quantum data types between modules. The syntax for defining a module is as follows:

{\footnotesize
\begin{alltt}
\textbf{return_type module} module_name
     (\textbf{type_1} parameter_1, ..., \textbf{type_n} parameter_n);
\end{alltt}
}

\noindent
Here, \verb+return_type+ can be null, integer, character, floating-point, double-precision floating-point, or structure. The argument list requirements are the same as those for the gate prototype function. To call the defined module, the following syntax is used:

{\footnotesize
\begin{alltt}
  module_name(parameter_1, ..., parameter_n);
\end{alltt}
}

\noindent
The overall structure of a Scaffold program, as depicted in Figure~\ref{fig:scaffold-structure}, is presented in~\cite{abhari2012scaffold}.

\end{document}